\documentclass[final,3p,times,twocolumn]{elsarticle}
\usepackage{amsmath,amsfonts}
\usepackage{amssymb}
\usepackage{algorithmic}
\usepackage{algorithm}
\usepackage{array}
\usepackage{textcomp}
\usepackage{url}
\usepackage{verbatim}
\usepackage{graphicx}
\usepackage{cite}
\usepackage{mathtools}
\usepackage{cases}
\usepackage{bigfoot}
\usepackage[table,xcdraw]{xcolor}
\usepackage{upgreek}
\usepackage{listings}
\usepackage{fancyvrb}
\usepackage[utf8]{inputenc}
\usepackage{pmboxdraw}
\usepackage{booktabs,caption}
\usepackage[flushleft]{threeparttable}
\usepackage{soul}

\usepackage{tikz}
\usepackage{pgfplots}
\usepackage{pgfplotstable}
\usetikzlibrary{babel} 
\usetikzlibrary{positioning,shadows}
\usetikzlibrary{arrows,shapes,automata,backgrounds,petri,calc,fadings,3d}
\usetikzlibrary{decorations.markings,decorations.pathmorphing,decorations.pathreplacing}
\usetikzlibrary{mindmap,trees}
\usetikzlibrary{calc}
\usepgfplotslibrary{groupplots,units,fillbetween}
\usepgfplotslibrary{fillbetween}

\usetikzlibrary{arrows,shapes,automata,backgrounds,petri,calc,fadings,3d}
\usetikzlibrary{decorations.markings,decorations.pathmorphing,decorations.pathreplacing}
\usepgfplotslibrary{groupplots,units,fillbetween,polar}
\usetikzlibrary{arrows.meta}
\tikzset{>=latex}
\usetikzlibrary{shapes.geometric, arrows}
\usetikzlibrary{arrows.meta}
\tikzset{>=latex}
\usetikzlibrary{calc}
\usepackage{makecell}%
\tikzstyle{startstop} = [rectangle, rounded corners, 
minimum width=3cm, 
minimum height=1cm,
text centered, 
draw=black, 
fill=red!30]
\pgfplotsset{compat=1.18}
\tikzstyle{io} = [trapezium, 
trapezium stretches=true, 
trapezium left angle=70, 
trapezium right angle=110, 
minimum width=3cm, 
minimum height=1cm, text centered, 
draw=black, fill=blue!10]

\tikzstyle{process} = [rectangle, 
minimum width=4cm, 
minimum height=1cm, 
text centered, 
text width=3cm, 
draw=black, 
fill=orange!30]

\tikzstyle{decision} = [diamond, 
minimum width=4cm, 
minimum height=2cm, 
text centered, 
draw=black, 
fill=green!30]
\tikzstyle{arrow} = [thick,->]

\definecolor{myazul}{rgb}{0,0.4431,0.7373}
\definecolor{myazul2}{rgb}{0,0.25,0.75}
\definecolor{mynaranja}{rgb}{0.8471,0.3216,0.0941}
\definecolor{myverde}{rgb}{0.47,0.67,0.19}
\definecolor{myverde2}{rgb}{0.0,0.5,0.0}
\definecolor{myvioleta}{rgb}{0.4,0.18,0.56}
\definecolor{myvioleta2}{rgb}{0.4,0.18,0.86}
\definecolor{myceleste}{rgb}{0.6,0.85,0.93}
\definecolor{myamarillo}{rgb}{0.93,0.88,0.17}
\definecolor{myrojo}{rgb}{0.84,0.04,0.00}
\definecolor{myrojo2}{rgb}{0.71,0 ,0}
\definecolor{mynegro}{rgb}{0.01,0.01,0.01}
\definecolor{backgroundColour}{rgb}{0.95,0.95,0.92}

\newcommand{\abs}[1]{\left|#1\right|}

\newcommand{\deff}{d_{\mathrm{eff}}}
\newcommand{\lcr}{L_{\mathrm{cr}}}

\newcommand{\derpar}[1]{\frac{\partial}{\partial #1}}
\newcommand{\derparn}[2]{\frac{\partial^{#1}}{\partial #2^{#1}}}

\newcommand{\dk}{\Delta k}

\newcommand{\fourier}[1]{\mathcal{F}\left\lbrace #1 \right\rbrace}
\newcommand{\invfourier}[1]{\mathcal{F}^{-1}\left\lbrace #1 \right\rbrace}

\newcommand{\Pin}{P_{\mathrm{in}}}

\newcommand{\etal}{\textit{et al.}}

\newcommand{\dispOp}[1]{\hat{\mathcal{D}}^{(\tau)}_{#1}}
\newcommand{\diffOp}[1]{\hat{\mathcal{D}}^{(xy)}_{#1}}
\newcommand{\conmut}[2]{[ #1, #2]}
\def\cpp{{C\nolinebreak[4]\hspace{-.05em}\raisebox{.4ex}{\tiny\bf ++}}}

\newcommand{\brahms}{\texttt{BRAHMS}}
\newcommand{\nx}{\mathrm{NX}}
\newcommand{\ny}{\mathrm{NY}}
\newcommand{\nz}{\mathrm{NZ}}
\newcommand{\nt}{\mathrm{NT}}
\newcommand{\n}{\mathrm{N}}

\definecolor{myred}{rgb}{0.71,0 ,0}
\newboolean{EDITEDCOLOR}
\setboolean{EDITEDCOLOR}{false} 
\newcommand{\edited}[1]{%
    \ifthenelse{\boolean{EDITEDCOLOR}}%
        {\textbf{\textcolor{myred}{#1}}}%
        {\textcolor{black}{#1}}%
}

\newcounter{bla}

\journal{Computer Physics Communications}

\begin{document}

\begin{frontmatter}



\title{\edited{BRAHMS: A cross-platform graphical toolkit for (3+1)D simulation of three-wave mixing in $\chi^{(2)}$ nonlinear media, with GPU and CPU backends}}

\author[a]{Alfredo Daniel Sanchez Rossi\corref{author}}
\author[b]{Daniel Alejandro La Valle Huignard}


\cortext[author] {Corresponding author.\\\textit{E-mail address:} alfredo.sanchez@tnuni.sk}

\address[a]{FunGlass - Centre for Functional and Surface Functionalized Glass, Alexander Dubček University of Trenčín, Slovakia}
\address[b]{Departament de Física de la Matèria Condensada, Universitat de Barcelona, C. Martí Franquès 1, Barcelona, 08028, Spain}


\begin{abstract}
\edited{We present \brahms, a cross-platform (Linux/Windows) graphical user interface, with GPU (CUDA) and CPU (OpenMP) backends, for the efficient and accurate simulation of three-wave mixing processes involving focused and pulsed Gaussian beams. The package solves the coupled $\chi^{(2)}$ nonlinear Schrödinger equations, including diffraction, dispersion, walk-off, and phase-mismatch effects simultaneously, in full (3+1)D. To our knowledge, this is the first open-source package that solves the complete second-order (three-wave-mixing) nonlinear-optics problem in (3+1)D while fully exploiting the parallel capabilities of modern GPUs, with an equivalent CPU backend available for users without GPU hardware. The GPU implementation is inherently scalable thanks to its \texttt{Thrust}-based design. The package provides a valuable tool for experimental design and for studying three-dimensional field propagation in nonlinear three-wave interactions, supporting applications such as second-harmonic generation (SHG), sum-frequency generation (SFG), and optical parametric generation (OPG). Its GUI requires no programming experience, so that users can readily set up simulations and interpret their results.}
\end{abstract}


\end{frontmatter}



{\bf PROGRAM SUMMARY/NEW VERSION PROGRAM SUMMARY}

\begin{small}
\noindent
{\em Program Title:~BRAHMS}                                          \\
{\em CPC Library link to program files:} (to be added by Technical Editor) \\
{\em Developer's repository link: https://github.com/alfredos84/BRAHMS} \\
{\em Code Ocean capsule:} (to be added by Technical Editor)\\
{\em Licensing provisions: MIT}  \\
{\em Programming language: \cpp17, CUDA, OpenMP, Python}                                   \\
{\em Supplementary material:}                                 \\
{\em Journal reference of previous version:}*                  \\
{\em Does the new version supersede the previous version?:}*   \\
{\em Reasons for the new version:*}\\
{\em Summary of revisions:}*\\
{\em Nature of problem: The problem solved in this work relates to typical three-wave mixing processes in a nonlinear crystal in a single-pass arrangement. While approaches to this problem exist, the novelty here lies in addressing the simultaneously spatial and temporal, (3+1)D, dependence of the electric fields. This aspect is typically disregarded in numerical applications due to memory constraints.
}\\
{\em Solution method: The coupled differential equations for three-wave interactions, which describe the field evolution along the crystal, are solved using the well-known Split-Step Fourier method (SSFM). This method involves the nonlinear crystal discretization into slices along the propagation direction of the fields (typically the $z$-direction) and calculating the linear and nonlinear effects separately. This, in turn, involves hundreds, or even thousands, of Fast Fourier Transform (FFT) calculations. Since our implementation considers both temporal and spatial effects, the FFTs are 1D and 2D (1D- and 2D-FFT, respectively). The coupled partial differential equations governing the propagation of electric fields along the nonlinear crystal are solved through two independent, physically equivalent backends: one written in \cpp/CUDA that leverages the parallel power of modern GPUs, and one written in \cpp/OpenMP for multi-core CPU execution, so that a GPU is not required to use the package. Both backends are accessed through BRAHMS, a cross-platform (Linux/Windows) GUI built in Python, which allows inexperienced users to set up, run, and post-process simulations that are typically 1D- or 3D-based without requiring programming knowledge.}\\

\end{small}

\section{Introduction}
\label{sec:intro}
\noindent
Nonlinear optical phenomena play a central role in modern photonics, enabling frequency conversion, ultrafast pulse shaping, spectroscopy, quantum light generation, and laser technology. Nonlinear processes such as second-harmonic generation (SHG), sum-frequency generation (SFG), and optical parametric generation (OPG) require modeling three-wave mixing (TWM) in second-order ($\chi^{(2)}$) media, which involves solving coupled nonlinear Schrödinger equations (NLSEs) that include transverse diffraction, group-velocity dispersion, phase mismatch, and nonlinear coupling between pump, signal, and idler fields~\citep{shen1984principles}. These processes are widely used in both fundamental research and industrial applications, ranging from frequency-agile laser sources to precision metrology and biomedical imaging.

Analytical treatments under plane-wave approximations provide valuable insight into conversion efficiency, phase-matching conditions, temporal walk-off, and group-velocity mismatch~\citep{arisholm1997general,Sanchez2022Ultrashort}, but fail to capture realistic effects arising, for instance, from spatial beam focusing~\citep{boyd1968parametric}. While such analytical approaches are extremely valuable for developing physical intuition, they rely on idealized assumptions that limit their predictive capability in realistic experimental configurations involving focused beams, finite apertures, or broadband excitation. Theoretical and numerical frameworks for modeling SHG of focused Gaussian beams in the continuous-wave (cw) regime, including thermal effects, have been proposed~\citep{sabouri2013thermal,sanchez2024cuda3D}. In the context of ultrashort pulses, the inclusion of dispersive effects becomes of paramount importance~\citep{pizzurro2024performance}. Several numerical studies have addressed TWM processes in optical parametric oscillators pumped in the cw or nanosecond regime~\citep{smith1995comparison, smith1999numerical, smith2003degenerate, sanchez2024cuda}. However, none of these approaches provides a unified framework capable of simultaneously incorporating both dispersion and diffraction, mainly due to the prohibitive memory requirements associated with their numerical implementation.

Generalized NLSE models have been derived to describe femtosecond pulse propagation under second-harmonic generation, including self-steepening, dispersion, diffraction, and mixed-derivative terms; nevertheless, their numerical treatment has typically been restricted to simplified geometries or effectively one-dimensional propagation~\citep{Trofimov2019}. More generally, numerical methods for systems of coupled NLSEs have been developed using finite-difference or time-splitting schemes, predominantly in the context of optical-fiber communications or other reduced-dimensional approximations~\citep{Ismail2016, MixedCNLS2002}. Although these methods are computationally efficient in lower dimensions, they do not directly extend to fully three-dimensional bulk configurations without severe computational penalties in memory footprint and execution time, which significantly limits their applicability to bulk nonlinear crystals under realistic focused-beam excitation. \edited{An alternative strategy, discussed further below, is to abandon the envelope (NLSE) description altogether and propagate the full electric field directly~\citep{gu2020simulation}.} More recently, the development of frequency-dependent and multifrequency nonlinear Schrödinger equations has further demonstrated that spatial confinement, spectral bandwidth, and nonlinear coupling can strongly influence wave-mixing dynamics, indicating that a comprehensive 3D+time solver is required for accurate modeling of ultrafast parametric processes~\citep{CastelloLurbe2024}. This gap is particularly relevant for researchers in computational physics and high-performance scientific computing, where scalable and memory-efficient algorithms are required to tackle large multidimensional nonlinear partial differential equation systems on modern accelerator-based architectures. Beyond nonlinear optics, the numerical strategy adopted here—namely, the large-scale operator-splitting integration of coupled dispersive nonlinear evolution equations on GPU architectures—may serve as a methodological reference for other (3+1)D systems governed by similar mathematical structures. Comparable challenges arise in plasma physics, Bose–Einstein condensate dynamics, nonlinear acoustics, and ultrafast matter-wave propagation, where the simultaneous treatment of multidimensional diffraction, dispersion, and nonlinear coupling leads to analogous computational demands.

\edited{To assess this gap systematically, Table~\ref{tab:comparison} compares~\brahms~against the closest available tools for numerically modeling second-order nonlinear wave-mixing and related coupled-NLSE problems, to the best of our knowledge. Widely used tools such as SNLO~\citep{smith2018crystal} provide fast, well-validated 1D/2D models (including focused-beam, undepleted- and depleted-pump SHG/OPO modules) but are CPU-only and do not solve the full (3+1)D problem with simultaneous diffraction and dispersion. General-purpose NLSE integrators such as NLSEmagic~\citep{caplan2013nlsemagic} offer GPU-accelerated, multi-dimensional solvers, but target the (typically single or coupled pairs of) scalar NLSE rather than the three-wave $\chi^{(2)}$ system with phase-mismatch and walk-off considered here. Beyond traditional numerical discretizations, data-driven approaches based on physics-informed neural networks combined with Fourier spectral methods have also been explored for related nonlinear Schr\"odinger-type systems, e.g.\ for the fractional NLS equation~\citep{wangyang2026fpinn}. The Maxwell-level, full-electric-field approach of Gu \etal~\citep{gu2020simulation}, introduced above, does solve the full (3+1)D problem, but at a substantially higher computational cost than the envelope (SSFM-based) description adopted by~\brahms. We are not aware of an existing open-source, GPU-accelerated package that solves the coupled $\chi^{(2)}$ NLSE system in full (3+1)D with simultaneous diffraction, dispersion, and walk-off for focused, pulsed Gaussian beams; commercial multiphysics/photonics suites may offer overlapping functionality but are closed-source and not benchmarked here.}

\begin{table}[h]
\centering 

\caption{Comparison of~\brahms~with related numerical tools for second-order nonlinear wave-mixing / coupled-NLSE problems.}
\resizebox{\linewidth}{!}{%
\label{tab:comparison}
\begin{threeparttable}
  \begin{tabular}{l|c|c|c}
    \toprule
    & Dimensionality & Diffraction + dispersion & GPU \\
    \midrule
    SNLO~\citep{smith2018crystal} & 1D / 2D+t & partial & -- \\
    NLSEmagic~\citep{caplan2013nlsemagic} & 1D/2D/3D & space only & \checkmark \\
    Gu \etal~\citep{gu2020simulation} & (3+1)D & \checkmark & -- \\
    \brahms~(this work) & (3+1)D & \checkmark & \checkmark \\
    \bottomrule
  \end{tabular}

\end{threeparttable}
}
\end{table}

\edited{In response to this need, we present~\brahms, a cross-platform (Linux/Windows) GUI, with GPU (CUDA) and CPU (OpenMP) backends, for the full (3+1)D coupled $\chi^{(2)}$ NLSE system, capable of simulating three-wave mixing with focused Gaussian beam inputs, including diffraction, dispersion, walk-off, and phase-mismatch effects. The GPU backend is recommended for large-scale (3+1)D simulations, enabling problem sizes that would otherwise be impractical on conventional CPU-only approaches. This tool aims to bridge the gap between simplified analytical theories and realistic large-scale numerical modeling of nonlinear frequency conversion in
bulk media.}

\edited{This paper is organized as follows. In Sec.~\ref{sec:theory}, the theoretical framework of the physical problem addressed by~\brahms~is presented. In Sec.~\ref{sec:numimpl}, the numerical implementation of the split-step Fourier method (SSFM) in (3+1)D is described. The software architecture and usage details are provided in Sec.~\ref{sec:softdesc}. Package performance and benchmarking is presented in Sec.~\ref{sec:performance}.
An illustrative (3+1)D simulation example is presented in Sec.~\ref{sec:examples}. Finally, conclusions are drawn in Sec.~\ref{sec:conclusions}.}


\section{Physical model}
\label{sec:theory}

The modeling of any three-wave mixing (TWM) processes consists of solving the so-called coupled wave equations (CWEs). These equations describe the evolution of electric fields along the spatial coordinate of propagation, within the nonlinear medium used. In this work we consider the more general case of forward propagation for the envelope of electric fields with space-time dependence $A(x,y,z,\tau)$. Each of the CWEs computes the evolution of an electric field at a specific wavelength, which interacts with the other two fields. In the three-wave parametric process, the interacting fields are pump, signal, and idler, each at angular frequencies, $\omega_p$, $\omega_s$, and $\omega_i$, respectively. For instance, in an optical parametric generation (OPG) process, a photon at $\omega_p$ is annihilated, and two photons at frequencies $\omega_s$ and $\omega_i$ are created. Conservation of energy dictates that $\omega_p=\omega_s+\omega_i$, while for parametric amplification and practical generation of coherent radiation at the signal and idler frequencies the phase-matching condition has to be satisfied, $\dk= 0$, where 
\begin{equation}\label{eq:mismatch}
    \dk=k_p-k_s-k_i,    
\end{equation}
is the mismatch factor, and $k_{\lambda}=n(\omega_{\lambda})\omega_{\lambda}/c$ ($x=p,~s,~i$) is the momentum at the wavelength, $\lambda$, with $n(\omega_{\lambda})$ the refractive index. The equations can generally be written as
\begin{numcases}{}
    \frac{\partial A_{p}}{\partial z} = i\kappa_p A_{s} A_{i}e^{-i\Delta k z} + \left(\dispOp{p}+\diffOp{p}\right)A_{p}, \label{eq:CEp}\\
    \frac{\partial A_{s}}{\partial z} = i\kappa_s A_{p} A_{i}^{*}e^{+i\Delta k z} + \left(\dispOp{s}+\diffOp{s}\right) A_{s}, \label{eq:CEs}\\
    \frac{\partial A_{i}}{\partial z} = i\kappa_i A_{p} A_{s}^{*}e^{+i\Delta k z} + \left(\dispOp{i}+\diffOp{i}\right) A_{i}, \label{eq:CEi}
\end{numcases}
where \edited{$(x,y,z)\in\mathcal{V}_{\text{crystal}}$ are the spatial coordinates inside the crystal}, $\kappa_{\lambda}=2\pi\deff/n_{\lambda}\lambda$ is the nonlinear coupling coefficient, and $\deff$ is the effective second-order susceptibility of the crystal. \edited{Here, $\dispOp{\lambda}$ are the temporal operators related to dispersion effects, whilst $\diffOp{\lambda}$ are the spatial operators related to diffraction and walk-off effects. These linear operators are defined as}
\begin{equation}\label{eq:dispOp}
    \dispOp{\lambda} = -\left[ \frac{\alpha_{\lambda}}{2}+ \left(\frac{1}{\nu_s} - \frac{1}{\nu_{\lambda}}\right) \frac{\partial}{\partial \tau}+i\frac{k^{''}_{\lambda}}{2}\frac{\partial^2}{\partial \tau^2} + i\frac{k^{'''}_{\lambda}}{3}\frac{\partial^3}{\partial \tau^3} \right],
\end{equation}
and 
\begin{equation}\label{eq:diffOp}
    \diffOp{\lambda} = -\frac{i}{2k_{\lambda}}\left(\derparn{2}{x}+\derparn{2}{y}\right) - \tan{\rho_{\lambda}\derpar{x}},
\end{equation}
Equation~\ref{eq:dispOp} indicates that the CWEs are written in a co-moving frame at the signal frequency ($\nu_s$) in the presence of linear attenuation, $\alpha_{\lambda}$, group-velocities, $\nu_{\lambda}=(\partial k/\partial\omega)^{-1}$, group-velocity dispersion (GVD), $k^{''}_{\lambda}=\partial^2 k/\partial\omega^2$, \edited{and third-order dispersion (TOD), $k^{'''}_{\lambda}=\partial^3 k/\partial\omega^3$}. The incorporation of higher-order dispersion terms beyond TOD is straightforward. \edited{The walk-off angle, $\rho_{\lambda}$, in Eq.~\ref{eq:diffOp} is relevant in birefringent nonlinear crystals. We note that the diffraction operator, $\diffOp{\lambda}$, is written under the paraxial approximation to the exact Helmholtz propagator, retaining only the leading-order transverse Laplacian term. This approximation is standard in SSFM treatments of beam propagation~\citep{powers2017fundamentals}.}

\subsection{Focused Gaussian beams}\label{sec:focusedgbs}
Typically, the experiments related to nonlinear processes discussed in this work involve Gaussian beams focused at the center of the nonlinear crystal. The idea of focusing arises from the need to increase the local intensity of the pump's electric field, enhancing the nonlinear effects. It is common practice to focus the pump beam at the center of the crystal, that is, at a distance $\lcr/2$. In the seminal paper of theory of focused Gaussian beams, an expression for SHG conversion efficiency as a function of the most relevant experimental parameters has been derived~\citep{boyd1968parametric}. The initial Gaussian pump beam is described by the electric field
\begin{equation}\label{eq:gaussianbeam}
A_p(x,y,z) = \frac{A_{p0}}{1+i\tau}\exp\left[-\frac{x^2+y^2}{w_{p}^2(1+i\tau)}+ik_{p}z\right]
\end{equation}
with 
$$ \tau = \frac{z-f}{z_R},~z_{R}=w_{\mathrm{F}}^2k_{\mathrm{F}}/2 , $$
where the subscripts $j=x,y$ are the transversal coordinates, $A_{\mathrm{F}0}$ is the electric field strength, $f$ and $w_{\mathrm{F}}$ are the focal points and the beam waists, and $z_{R}$ is the Rayleigh range. Equation~\ref{eq:gaussianbeam} describes a Gaussian beam propagating along the $z$ direction. For a beam focused at the center of the crystal, $f = \lcr/2$, the initial electric field at the entrance of the nonlinear crystal is
\begin{equation}\label{eq:gaussianbeamc}
A_p(x,y,z=0) = \frac{A_{p0}}{1-i\xi}\exp\left[-\frac{x^2+y^2}{w_{p}^2(1-i\xi)}\right],
\end{equation}
where the \textit{focusing parameter}, defined as $\xi=\lcr/2z_{R}$, is widely used in the literature~\citep{kumar2009high,sabouri2013thermal}. The focusing parameter is of crucial practical importance, as it determines the conversion efficiency in SHG experiments. We set the initial electric field at the SHG wavelength as Gaussian white noise with both random amplitude and phase.

\section{Numerical algorithm}\label{sec:numimpl}

Among the number of available algorithms to model the evolution of electric fields in dielectric media, the SSFM is widely used for three-~\citep{arisholm1997general} and four-wave mixing problems in the realm of nonlinear optics~\citep{agrawal2000nonlinear,Sinkin03}.

\subsection{Split-step Fourier method (SSFM)}
\label{sec:ssfm}
\noindent

\begin{figure*}
    \centering
    \includegraphics[width=0.9\linewidth]{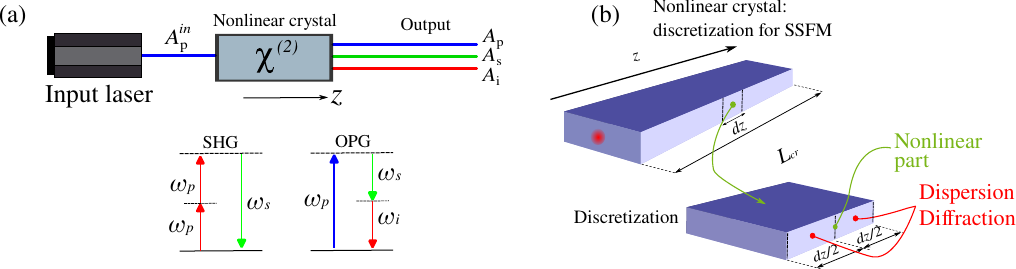}
    \caption{Schematic of the system simulated by~\brahms. (a) Typical TWM experiment: an input pump laser enters in a second-order nonlinear medium to generate new frequencies. (b) Nonlinear discretization along $z$-direction, in which linear and nonlinear effects are considered.}
    \label{fig:scheme}
\end{figure*}

The SSFM is used here to model the propagation along the nonlinear medium, as schematically shown in Fig.~\ref{fig:scheme}.
The crystal with length, $\lcr$, is discretized along the $z-$direction into steps of length, $dz$. In every step, the SSFM simultaneously solves the linear and nonlinear effects. The linear part is solved in the frequency domain, and the nonlinear part is solved in the time domain using a four-order Runge-Kutta method. This sequence is repeated throughout the length of the crystal. Depending on the implementation, this algorithm exhibits an error, $\mathcal{O}(dz^2)$ or $\mathcal{O}(dz^3)$. This is sequentially solved along the entire crystal and requires many operations with complex vectors as well as discrete Fourier transforms (DFTs) during the simulation. Equations~\ref{eq:CEp},~\ref{eq:CEs}, and~\ref{eq:CEi} can be written in the matrix form as (omitting the superscripts $m$)
\begin{equation}\label{eq:CWEsmatrix}
    \frac{\partial}{\partial z} 
    \begin{pmatrix}
    A_p\\
    A_s\\
    A_i
    \end{pmatrix} = 
    \underbrace{\begin{pmatrix}
    \hat{L}_p & 0 & 0  \\
    0 & \hat{L}_s & 0  \\
    0 & 0 & \hat{L}_i 
    \end{pmatrix}}_{\text{Linear operator}\\ \hat{L}}   
    \begin{pmatrix}
    A_p\\
    A_s\\
    A_i
    \end{pmatrix}
    +
    \underbrace{\begin{pmatrix}
    0 & 0 & \hat{N}_p \\
    \hat{N}_s & 0 & 0 \\
    \hat{N}_i & 0 & 0  
    \end{pmatrix}}_{\text{Nonlinear operator~}\\ \hat{N}}   
    \begin{pmatrix}
    A_p\\
    A_s\\
    A_i
    \end{pmatrix}
\end{equation}
where the linear operators $\hat{L}_{\lambda}=\dispOp{\lambda}+\diffOp{\lambda}$ are given by Eqs.~\ref{eq:dispOp} and \ref{eq:diffOp}, and $\hat{N}_{\lambda}$ are the corresponding nonlinear operators for each wavelength, $\hat{N}_p = i\kappa_p A_s e^{-i\dk z}$, $\hat{N}_s = i\kappa_s A_i^* e^{i\dk z}$, and  $\hat{N}_i = i\kappa_i A_s^* e^{i\dk z}$, respectively. Equation~\ref{eq:CWEsmatrix} is then reduced to
\begin{equation}
    \frac{\partial \Vec{A}}{\partial z}  = \left( \hat{L} + \hat{N} \right) \Vec{A}
\end{equation}
with a symbolic solution given by
\begin{equation}\label{eq:evol}
    \Vec{A}(z+dz) = e^{\left( \hat{L} + \hat{N} \right)dz} \Vec{A}(z).
\end{equation}
Since the operators, $\hat{L}$ and $\hat{N}$, in general do not commute, the approximation $$e^{\left( \hat{L} + \hat{N} \right)dz} \approx e^{ \hat{L}dz} e^{\hat{N} dz},$$
that yields an error, $\mathcal{O}(dz^2)$, is often used. However, in this work we implement a more accurate expression~\citep{agrawal2000nonlinear}
\begin{equation}\label{eq:expon}
    e^{\left( \hat{L} + \hat{N} \right)dz} \approx e^{ \hat{N}\frac{dz}{2}}e^{ \hat{L}dz} e^{ \hat{N}\frac{dz}{2}},
\end{equation}
with an error of $\mathcal{O}(dz^3)$. In this scheme, every step is solved by computing the nonlinear term in the first half-step, $dz/2$. After one Fourier transform, the linear term is computed in the entire step, $dz$. Finally, the nonlinear term is again computed in the second half-step, $dz/2$. This sequence, $\hat{N}/2-\hat{L}-\hat{N}/2$, is equivalent to its counterpart, $\hat{L}/2-\hat{N}-\hat{L}/2$, since both lead to the same solution. By inserting Eq.~\ref{eq:expon} in Eq.~\ref{eq:evol} and solving the linear part in the frequency domain, the field evolution reads
\begin{equation}
    \Vec{A}(z+dz) \approx e^{ \hat{N}\frac{dz}{2}} \invfourier{e^{\hat{L}dz} \fourier{e^{\hat{N}\frac{dz}{2}}\Vec{A}(z)}} ,
\end{equation}
where $\fourier{\cdot}$ stands for the Fourier transform. Finally, it is worth noting that if the electric fields are smooth enough, the equality of mixed partial derivatives follows from Schwarz's theorem, guaranteeing that the dispersion and diffraction operators in Eqs.~\ref{eq:dispOp} and \ref{eq:diffOp} commutes: $\conmut{\dispOp{\lambda}}{\diffOp{\lambda}}=0$.

\subsection{Vector size and grid discretization considerations}
\label{sec:vecsizeconsid}
\noindent

The choice of the number of vector elements, as well as the step size of the spatial and temporal-spectral grids will depend on each specific problem. Here we simply provide a guide on what to consider when tackling the problem to be modeled.
\edited{The criteria for choosing the spatial grid that we will give is based on two basic aspects related to the stability of the algorithm~\citep{sanchez2024cuda3D}. As described in Section~\ref{sec:ssfm}, the numerical error of this algorithm is $\mathcal{O}(dz^3)$, so decreasing the step size, $dz$,  the approximation of the numerical solution will be closer to the analytical solution. One of the problems that can arise when the value of $dz$ is not small enough is that the code returns \texttt{NaN} values. This can be seen in the nonlinear part of the coupled equations. For simplicity, consider the case of the dispersionless SHG process with perfect phase-matching ($\Delta k = 0$) and focus on the equation for the signal that can be approximated by}

$$\frac{\partial A_{s}}{\partial z} = i\kappa_s A_{p} A_{s}^{*} \Rightarrow A_s(z+dz) \approx A_s(z) + i \Delta A_s(z),$$

\edited{where $\Delta A_s(z) = \left(\kappa_s A_p(z) dz \right) A_s^*(z)$ is the incremental change in the signal electric field, with $\kappa_s = 2\pi \deff/n_s\lambda_s$. With this approach, we expect the change in the electric field to be incremental, that is}
\begin{equation*}
\abs{\Delta A_s(z)} \ll \abs{A_s(z)} \approx \abs{A_s(z+dz)},    
\end{equation*}
\edited{in order to avoid undesired drastic changes. Since $\kappa_s$ and $A_p$ are experimental parameters, the only way to prevent any divergence is by varying $dz$. For example, for high pump powers or for crystals with a high nonlinear coefficient, $\deff$, it will be necessary to pay attention to the set value of $dz$~\citep{smith1999numerical}.}\\

\section{Package~\brahms}
\label{sec:softdesc}

\brahms~is a simulation package whose computational core is provided by \edited{two independent, physically equivalent backends: a} \cpp/CUDA \edited{backend, which} makes extensive use of the \texttt{Thrust} library, a CUDA-compatible parallel STL\edited{, and depends on the \texttt{cuda-toolkits}~\citep{NVT}, and a \cpp/OpenMP backend for multi-core CPU execution that requires no GPU hardware at all. Both backends solve the CWEs and SSFM described in Secs.~\ref{sec:theory} and~\ref{sec:numimpl}, and produce numerically equivalent results (up to floating-point round-off) for the same input configuration, which we use as a cross-validation check between the two independent implementations. The GPU backend remains the recommended choice for large-scale problems, as it} exhibits a high degree of inherent data parallelism, making it particularly well suited for GPU-accelerated implementations. By relying on \texttt{Thrust}, the implementation avoids much of the repetitive low-level CUDA code, resulting in a cleaner, more maintainable, and robust codebase. This high-level parallel abstraction also improves scalability, allowing the solver to naturally extend to larger spatial grids, longer propagation distances, and more demanding nonlinear optical configurations. \edited{Both backends are accessed through a single cross-platform GUI written in Python, so that no programming knowledge is required to set up, run, and post-process simulations. The package was tested on both Linux and Windows systems.}

\subsection{Backend \cpp/CUDA}

\edited{The package backend has a structured folder layout as follows:}
\begin{Verbatim}
engine_gpu/
├── twm.cu
└── headers/
    ├── Common.cuh
    ├── Crystal.cuh
    ├── DataTypes.cuh
    ├── EFields.cuh
    ├── Files.cuh
    ├── Libraries.cuh
    ├── Operators.cuh
    ├── PackageLibraries.cuh
    ├── PhaseMatching.cuh
    ├── SaveOutputs.cuh
    ├── Solver.cuh
    └── Tfield.cuh
 \end{Verbatim}

In the code, the electric fields, the nonlinear crystals and the solver of differential equations are objects belonging to their corresponding classes that interact through their methods throughout the simulation.
The package contains header files in the folder named~\texttt{headers}, which can be either modified or adapted to the user specific applications. The following is a summarized description of the header files:
\begin{itemize}
    \item \texttt{Common.cuh} contains CUDA error-checking macros and small utility functions.
    \item \edited{\texttt{Crystal.cuh} contains the \texttt{class Crystal}, populated at runtime from the JSON configuration file rather than from per-material hard-coded subclasses; any crystal defined through the GUI's crystal database (Sec.~\ref{sec:pyGUI}) can therefore be used without modifying or recompiling the source code. Both quasi-phase-matched and birefringent nonlinear crystals are supported.}
    \item \texttt{DataTypes.cuh} contains all relevant data-type declarations and constants. In the package, single-precision data types are used for real and complex scalars and matrices:
    \begin{lstlisting}[language=C++, basicstyle=\ttfamily, showstringspaces=false,
  commentstyle=\color{green!50!black},
  keywordstyle=\color{blue},
  basicstyle=\ttfamily\footnotesize,]
// using...
// for real scalars
real_t = float;
// for complex scalars
complex_t = cufftComplex;
// for real matrices in host
rVech_t = thrust::host_vector<real_t>;
// for real matrices in device
rVecd_t = thrust::device_vector<real_t>;
// for complex matrices in host
cVech_t = thrust::host_vector<complex_t>;
// for complex matrices in device
cVecd_t = thrust::device_vector<complex_t>;
// for json files
json = nlohmann::json;
    \end{lstlisting}
    \item \texttt{EFields.cuh} contains the definition of the \texttt{class EFields} and its methods.
    \item \texttt{Files.cuh} contains functions useful to save real or complex vectors, matrices and tensors into a \texttt{.dat} file.
    \item \texttt{Libraries.cuh} centralizes the standard-library includes used throughout the package.
    \item \texttt{Operators.cuh} provides overloaded arithmetic operators (\(+,-,*,/\)) that enable transparent manipulation of real and complex numbers as well as electric fields represented as \texttt{Thrust} vectors.
    \item \edited{\texttt{PackageLibraries.cuh} is a single entry-point header that fixes the include order of all the files above.}
    \item \edited{\texttt{PhaseMatching.cuh} manages the (possibly spatially varying) phase mismatch $\dk(x,y,z)$, including its temperature dependence for QPM crystals when the thermal model (\texttt{Tfield.cuh}, below) is enabled.}
    \item \texttt{SaveOutputs.cuh} implements HDF5 output routines for storing electric fields and associated time and frequency vectors produced by the simulations.
    \item \texttt{Solver.cuh} contains the definition of the \texttt{class Solver} and its methods. This library contains the SSFM routines.
    \item \edited{\texttt{Tfield.cuh} solves the steady-state heat equation, $\nabla^2 T=-Q/\kappa$, on the same spatial grid as the optical fields, for the optional thermal model.}
\end{itemize}
\edited{An entirely analogous header layout implements the same class structure for the \cpp/OpenMP backend, replacing CUDA kernels and \texttt{Thrust} device vectors with OpenMP-parallel loops and standard host arrays, and \texttt{cuFFT} with \texttt{FFTW3}~\citep{frigo2005design} (Sec.~\ref{sec:cpubackend}).}

\subsection{Python GUI}
\label{sec:pyGUI}
\noindent
\edited{User interaction with both backends takes place entirely through~\brahms's GUI, organized into a small number of tabs that mirror the typical workflow of a three-wave-mixing experiment:}
\begin{itemize}
    \item \edited{\textbf{Nonlinear crystals.} Users define the nonlinear crystal directly from its Sellmeier equation(s), specifying whether the crystal is quasi-phase-matched (QPM) or birefringent, together with the relevant material parameters ($\deff$, linear and two-photon absorption, walk-off angle, thermal properties, grating period for QPM crystals, etc.). A reduced set of common crystals (e.g., $\upbeta$-BaB$_2$O$_4$ (BBO), MgO:PPLN, MgO:sPPLT) is pre-loaded with literature Sellmeier coefficients, and users can add, edit, or remove their custom crystals.}
    \item \edited{\textbf{Phase matching.} Given the selected crystal and a set of experimental parameters (pump wavelength, temperature, polarization, and, for QPM crystals, grating period), this tab computes the phase-matching conditions ($\dk$ as a function of the relevant parameter) for the chosen three-wave-mixing process (SHG, SFG, or OPG).}
    \item \edited{\textbf{Single Simulation.} This tab runs a single (3+1)D or reduced-dimensionality simulation for a given configuration, on either the CPU or the GPU backend, as selected by the user.}
    \item \edited{\textbf{Parameter sweep.} This tab automates repeated simulations while scanning one physical parameter (e.g., pump power, beam waist, crystal temperature, or phase mismatch) to compute conversion-efficiency curves and identify optimal operating conditions, such as the beam waist or temperature that maximizes conversion efficiency.}
    \item \edited{\textbf{(3+1)D simulation.} This tab runs the full space- and time-resolved simulation in the \texttt{focused-pulsed} mode (Sec.~\ref{sec:examples}), the most demanding configuration supported by the package. Because of its computational cost, the use of the GPU backend is strongly recommended for this mode in order to obtain results within a practical runtime.}
\end{itemize}
\edited{Behind the GUI, each backend is invoked exactly as in a command-line workflow: the physical and grid parameters selected by the user are written to a JSON \textit{configuration file}, the appropriate backend binary is compiled (if not already available for the requested grid size) and executed, and the resulting HDF5 output is read back and post-processed for on-screen visualization. This keeps the underlying, reproducible, file-based workflow intact while removing the need for users to write or edit any code.}

\subsection{Algorithmic flowchart and implementation considerations}

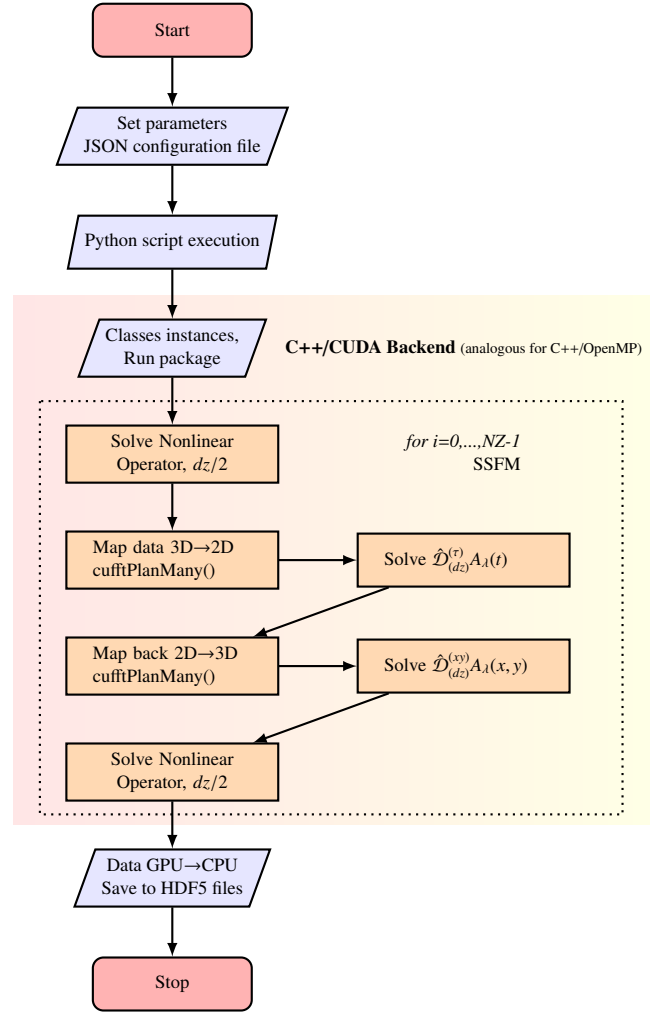
\begin{figure}[t]
\centering
\begin{tikzpicture}[node distance=2cm,thick,scale=0.7, every node/.style={transform shape},->]
\shade[black, dotted, thin, left color=red!10, right color=yellow!10] (-3,-5) rectangle (9,-15);
\node[font=\normalsize] at (5.5,-6) {\textbf{C++/CUDA Backend} \footnotesize(analogous for C++/OpenMP)};
\draw[black, dotted] (-2.5,-7) rectangle (8.5,-14.8);
\node[font=\normalsize] at (5.5,-8) {\makecell[r]{\textit{for i=0,...,NZ-1}\\SSFM}};
\node (start) [startstop] {Start};
\node (in1) [io, below of=start] {\makecell[c]{Set parameters\\JSON configuration file}};
\node (in12) [io, below of=in1] {Python script execution};
\node (in2) [io, below of=in12] {\makecell[c]{Classes instances,\\Run package}};
\node (proc1) [process, below of=in2] {\makecell[c]{Solve Nonlinear\\Operator, $dz/2$}};
\node (proc2) [process, below of=proc1] {\makecell[l]{Map data 3D$\rightarrow$2D\\cufftPlanMany()}};
\node (proc3) [process, right of=proc2, xshift=3.5cm] {\makecell[l]{Solve $\dispOp{(dz)}A_\lambda(t)$}};
\node (proc4) [process, below of=proc2] {\makecell[l]{Map back 2D$\rightarrow$3D\\cufftPlanMany()}};
\node (proc5) [process, right of=proc4, xshift=3.5cm] {\makecell[l]{Solve $\diffOp{(dz)}A_\lambda(x,y)$}};
\node (proc6) [process, below of=proc4] {\makecell[c]{Solve Nonlinear\\Operator, $dz/2$}};
\node (out1) [io, below of=proc6] {\makecell[c]{Data GPU$\rightarrow$CPU\\Save to HDF5 files}};
\node (stop) [startstop, below of=out1] {Stop};
\draw [arrow] (proc2) --  (proc3);
\draw [arrow] (proc4) --  (proc5);
\draw [arrow] (proc3) --  (proc4);
\draw [arrow] (proc5) --  (proc6);
\draw [arrow] (start) --  (in1);
\draw [arrow] (in1) --  (in12);
\draw [arrow] (in12) --  (in2);
\draw [arrow] (in2) --  (proc1);
\draw [arrow] (proc1) --  (proc2);
\draw [arrow] (proc6) --  (out1);
\draw [arrow] (out1) --  (stop);
\end{tikzpicture}
\caption{Flowchart package}
\label{fig:flowchart}
\end{figure}

Figure~\ref{fig:flowchart} shows the flowchart of the implementation adopted by our package for an execution in (3+1)D scheme\edited{, using the \cpp/CUDA backend as illustration; the \cpp/OpenMP backend follows the same sequence of steps, with CUDA kernel launches and \texttt{Thrust} device vectors replaced by OpenMP-parallel loops and host arrays (Sec.~\ref{sec:cpubackend})}. All variables and objects generated in the execution of the code are declared as being on the GPU\edited{, in this case}. Furthermore, the entire code is executed on the GPU \edited{(or, for the CPU backend, in parallel across the available OpenMP threads)}.

As can be seen in Fig.~\ref{fig:flowchart}, the SSFM solves the CWEs in the time, spectral and spatial domains for the (3+1)D case. In our implementation, the electric fields are \texttt{Thrust} vectors indexed in the coordinates ($x,y,\tau$) that evolve inside the crystal along the coordinate $z$. That is, for each position in $z$, all tensors are updated as the SSFM progresses. In this implementation, the electric fields within each position in $z$ are not stored for memory reasons. The user could easily add routines to allow this, although it is not recommended. Since the vector size is NX$\times$NY$\times$NT, and for complex scalars this represent a twofold memory requirement, hence the memory size might be limited to NX=NY=2$^{7}$ and NT=2$^{11}$ or 2$^{12}$ grid points, depending on the available GPU. \edited{We emphasize that this is a hardware limitation, not an algorithmic one. Since the field at each $z$-slice is not retained, the memory footprint is independent of $\nz$ and scales only with the transverse-temporal grid $\n=\nx\times\ny\times\nt$. Counting all single-precision complex-valued working arrays required by the solver gives approximately $37$ arrays of size $\n$, i.e., a memory requirement of $\approx 37\times 8\,\mathrm{B}\times \n \approx 300\,\mathrm{B}\times\n$. For the largest grid reported in Table~\ref{tab:gpu_runtime} ($\nx=\ny=128$, $\nt=4096$, $\n\approx 6.7\times 10^{7}$), this amounts to $\approx 20$~GB. Larger transverse-temporal grids are therefore directly accessible on higher-VRAM cards, with no algorithmic modification required. Multi-GPU domain decomposition, which would relax this limit further by distributing the transverse grid across devices, is not implemented in the present version and is left for future work.}

All \texttt{Thrust} vectors in the package are indexed through the \texttt{IDX(x,y,$\tau$)} function defined in \texttt{Common.cuh}. While correct indexing is essential for performing element-wise operations, it is equally important to understand how this indexing scheme distributes data in memory. According to the definition of \texttt{IDX(x,y,$\tau$)}, a sequential sweep through the vector establishes the hierarchy $x \rightarrow y \rightarrow \tau$, meaning that $x \in [0,\mathrm{NX}-1]$ varies fastest, followed by $y \in [0,\mathrm{NY}-1]$, and finally $\tau \in [0,\mathrm{NT}-1]$. Under this ordering, the data layout is naturally suited for evaluating the diffractive operator in Eq.~\ref{eq:diffOp}, since it allows the use of CUDA’s batched FFT functionality to perform NT independent 2D FFTs, one for each temporal slice, $\tau$.

\begin{figure*}
    \centering
    \includegraphics[width=0.8\linewidth]{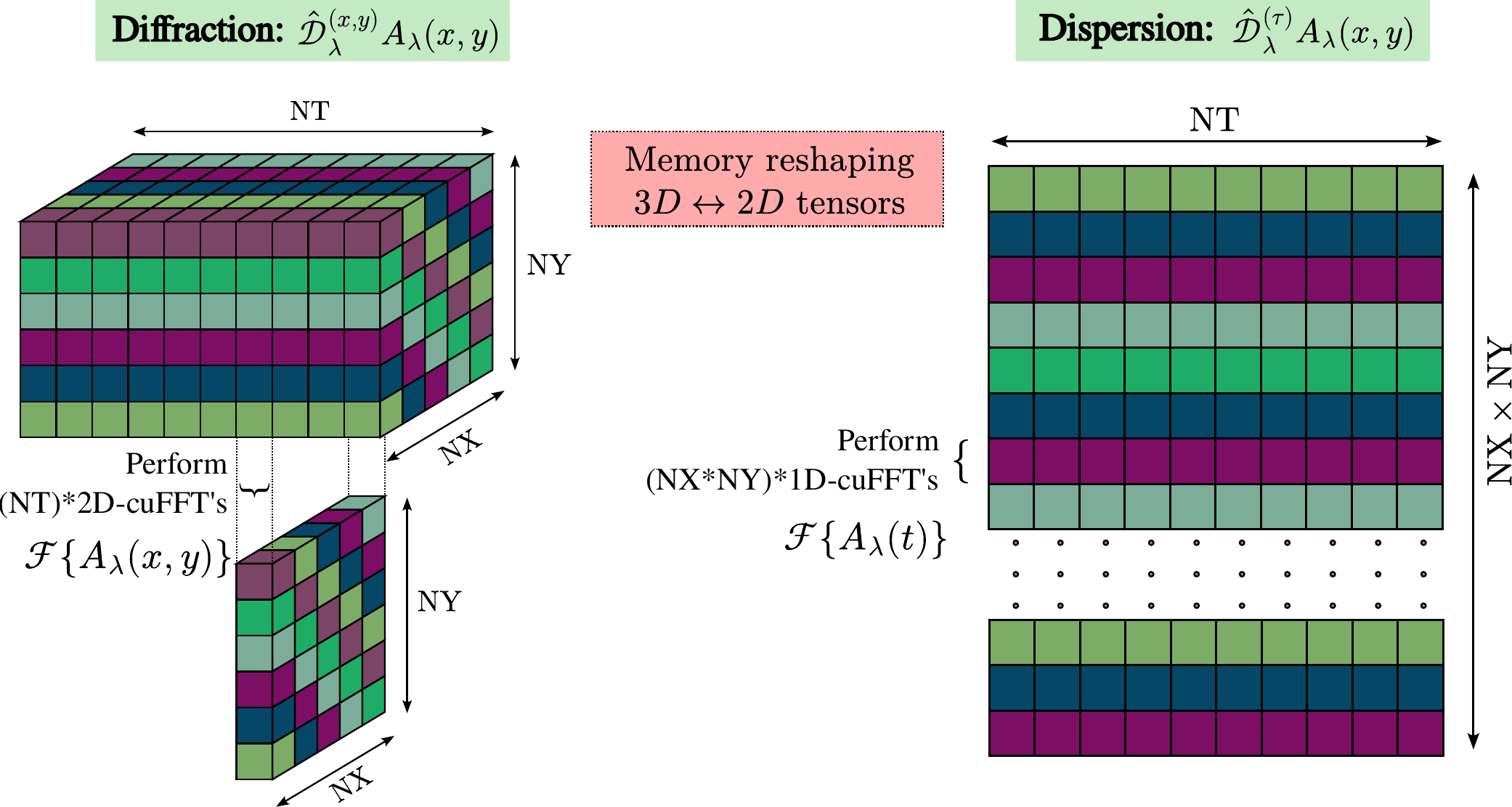}
    \caption{Memory reshaping for computing diffraction and dispersion terms. (left side) The diffraction term is calculated using a memory structure according to the index hierarchy $x\rightarrow y\rightarrow\tau$, in which 2D FFT are performed for each $\tau$, taken in batches of $\nx\times\ny$ elements (see the bottom of the tensor). (right side) For the calculation of the scattering term, the index hierarchy $\tau\rightarrow \textrm{pair}{(x,y)}$ is restructured, in which each row of the new configuration corresponds to a vector of times located in the $\textrm{pair}{(x,y)}$. Consequently, $\nx\times\ny$ 1D-FFTs are performed in the time domain.}
    \label{fig:memoryreshape}
\end{figure*}

However, this layout is not appropriate for evaluating the dispersive term in Eq.~\ref{eq:dispOp}, which requires performing NX×NY independent 1D FFTs along the temporal dimension. For this reason, an intermediate memory-reordering step was implemented. The reordering maps the 3D array into a 2D layout where elements sharing the same $(x,y)$ index become contiguous, enabling an efficient batch of $\nx\times\ny$ 1D FFTs along $\tau$. This prevents severe performance bottlenecks that would otherwise arise from operating on non-contiguous memory segments. Figure~\ref{fig:memoryreshape} illustrates the memory transformation used for both diffractive and dispersive calculations. This memory management is the key step of our efficient implementation.

\section{Package performance}
\label{sec:performance}
\noindent

The package essentially performs sums, products and discrete 1D- and 2D-Fourier transforms (1D- and 2D-DFTs, respectively) of real- and complex-valued tensors. Since the package does not store the fields at each position within the crystal when the time dependency is considered, i.e. $\nt>1$, the required (2+1)D grid $(x,y,\tau)$ size is: $\n =\nx\times \ny \times \nt $. For users who need to store fields at specific points within the nonlinear crystal, the \texttt{Solver::SSFM()} function in the package needs to be accordingly modified by adding the corresponding functions stored in \texttt{Files.cuh} library. DFTs were calculated using \texttt{cuFFT}, the CUDA library for computing Fourier transforms. Both 1D- and 2D-DFTs implementations have an order of convergence, $\mathcal{O}\left(n\log(n)\right)$, with $n$ the involved number of points~\citep{frigo2005design, cuFFT}. On the other hand, the rest of operations of sums and products have an order of convergence, $\mathcal{O}(n)$. The global algorithm has a convergence order dominated by the DFT. 
Table~\ref{tab:gpu_runtime} shows the execution times of (3+1)D simulations for different grid sizes and different GPU cards. 
\begin{table}[h]
\centering
  \begin{threeparttable}
    \caption{Simulation Runtime in seconds in pulsed-focused mode for different GPUs and different grid sizes.}
     \begin{tabular}{c|c|c|c|c|c}
        \toprule
        N & $\nx$ & $\nt$ & GPU-1 & GPU-2 & GPU-3  \\
        \midrule
        $2^{20}$ & 64 & 256 & 843 & 722 & 1032  \\
        $2^{21}$ & 64 & 512 & 1761 & 1573 & 2059 \\
        $2^{22}$ & 64 & 1024 & 3654 & 3211 & 4097 \\
        $2^{23}$ & 64 & 2048 & 7390 & 6444 & 8240 \\
        $2^{24}$ & 64 & 4096 & 14791 & 12924 & 16611 \\     
        \midrule
        $2^{22}$ & 128 & 256 & 3513 & 2681 & 1296  \\
        $2^{23}$ & 128 & 512 & 6681 & 5213 & 2576 \\
        $2^{24}$ & 128 & 1024 & 13899 & 10486 & 5141 \\
        $2^{25}$ & 128 & 2048 & - & 21037 & 10305 \\
        $2^{26}$ & 128 & 4096 & - & 42207 & 20698 \\ 
        \bottomrule
     \end{tabular}\label{tab:gpu_runtime}
    \begin{tablenotes}
      \footnotesize
      \item[-] List of GPUs NVIDIA used in this work. GPU-1: GeForce RTX 4070 Laptop ; GPU-2: RTX 4000 ADA; GPU-3: A100 \edited{40 GB}.
      \item[-] Data with '--' represent simulations that could not be performed due to      capacity limitations.
      \item[-] In all the simulations $z$-discretization is $\nz=100000$ and $\nx=\ny$.
    \end{tablenotes}
  \end{threeparttable}
\end{table}

\subsection{CPU (OpenMP) backend and CPU/GPU performance comparison}
\label{sec:cpubackend}
\noindent

\edited{While~\brahms~was originally designed to run on GPU, and this remains the recommended execution mode for large-scale (3+1)D simulations, we have implemented an equivalent CPU-only backend so that the package can also be used on machines without an NVIDIA GPU.}

\subsubsection{Implementation}

\edited{The CPU backend solves the identical physical model and numerical algorithm described in Secs.~\ref{sec:theory} and~\ref{sec:numimpl}, re-implemented in \cpp17 with OpenMP
multi-threading in place of CUDA/\texttt{Thrust}. The main implementation
differences are:}
\begin{itemize}
    \item \edited{\textbf{Fourier transforms}. Batched 1D- and 2D-DFTs are computed with \texttt{FFTW3}~\citep{frigo2005design} (single precision, multi-threaded, \texttt{FFTW\_MEASURE} planning) instead of \texttt{cuFFT}. The same memory-reshaping strategy of Fig.~\ref{fig:memoryreshape} is used, so that both the dispersive and diffractive FFT batches operate on contiguous memory.}
    \item \edited{\textbf{Parallelization}. Element-wise operations (nonlinear step, propagator products, memory reshaping) are parallelized across the transverse grid with \texttt{\#pragma omp parallel for}, instead of CUDA thread blocks.
    \item \textbf{Precision and physical model.} Identical to the GPU backend: single-precision real and complex data types, and the same dispersion, diffraction, absorption, two-photon-absorption, and walk-off terms are supported (Sec.~\ref{sec:theory}).}
\end{itemize}
\edited{No changes to the physical model or the JSON configuration schema are required to switch between the GPU and CPU backends; the same input file produces physically identical results (up to floating-point round-off) on either backend, which we use as a cross-validation check between the two independent implementations.}

\subsubsection{Benchmark setup and strong scaling}

\edited{CPU benchmarks were obtained on an Intel Core i7-10510U (4 physical cores, 8 logical threads via Hyper-Threading, 8~MB L3 cache, DDR4 dual-channel memory), compiled with \texttt{g++ -O3 -march=native -fopenmp}. This is a modest, consumer-grade laptop CPU, chosen deliberately as a conservative baseline: any speedup reported below would be larger still against a lower-end CPU, and the qualitative trends are expected to hold (with different absolute numbers) on higher-end multi-core server CPUs. All simulations use the same \texttt{pulsed-focused} configuration and grid sizes as Table~\ref{tab:gpu_runtime}, on the same MgO:PPLN crystal ($\lcr=20$~mm), with $\nz=100\,000$.}

\edited{Table~\ref{tab:cpu_scaling} shows the resulting \emph{strong scaling} of the CPU backend --- runtime as a function of the number of OpenMP threads at fixed problem size (as opposed to \emph{weak scaling}, where the problem size grows in proportion to the thread count). The 4-thread configuration (matching the 4 physical cores of the test CPU) achieves the best performance, $1.60\times$ over a single thread; enabling the additional 4 Hyper-Threading logical cores (4$\to$8 threads) \emph{degrades} performance by $\sim$27\% relative to 4 threads. This is expected for a memory-bandwidth-bound, FFT-dominated workload: paired logical cores share L1/L2 cache and execution units without providing additional memory bandwidth, so Hyper-Threading does not help once the physical cores are already bandwidth-saturated.}

\begin{table}[h]
\centering
\begin{threeparttable}
  \caption{Strong scaling of the CPU (OpenMP) backend. $\nx=\ny=64$, $\nt=512$, $\nz=100\,000$.}
  \begin{tabular}{c|c|c}
    \toprule
    OpenMP threads & $t$ (h) & Speedup vs.\ 1 thread \\
    \midrule
    1 & 28.2 & $1.00\times$ \\
    2 & 20.4 & $1.38\times$ \\
    4 & 17.6 & $1.60\times$ \\
    8 & 24.1 & $1.17\times$ \\
    \bottomrule
  \end{tabular}\label{tab:cpu_scaling}
  \begin{tablenotes}
    \footnotesize
    \item[-] Test CPU: Intel Core i7-10510U (4 physical cores / 8 logical threads).
  \end{tablenotes}
\end{threeparttable}
\end{table}

\subsubsection{CPU vs.\ GPU speedup for (3+1)D simulations}

\begin{figure*}[h]
    \centering
    \includegraphics[width=\linewidth]{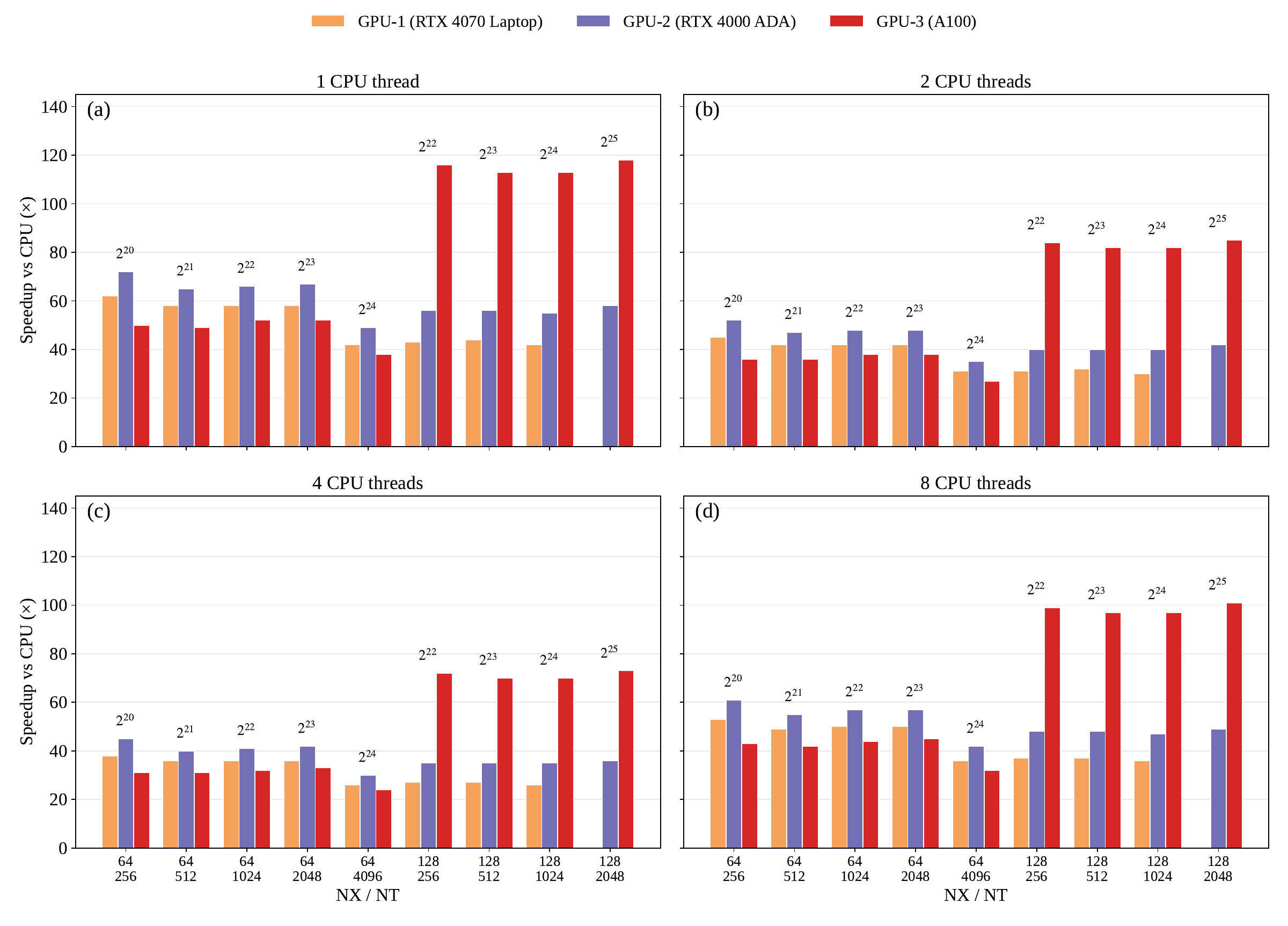}
    \caption{CPU (8 OpenMP threads) vs.\ GPU speedup, as  $=T_{\mathrm{CPU}}/T_{\mathrm{GPU}}$. GPU-1: RTX 4070 Laptop; GPU-2: RTX 4000 ADA; GPU-3: A100 (same cards as Table~\ref{tab:gpu_runtime}). Missing bars denote cases that could not be run due to capacity limitations on the respective hardware. $\nz=100\,000$, $\nx=\ny$ in all cases, as in Table~\ref{tab:gpu_runtime}.}
    \label{fig:speedup}
\end{figure*}

\edited{The resulting speedup measurements for each of the three GPUs respect to the CPU implementation are depicted in Fig.~\ref{fig:speedup}. Two trends are apparent. First, the GPU speedup is substantial across the whole grid range, $> 30\times$ depending on grid size and GPU model; even the consumer-grade laptop GPU (GPU-1) turns a 12-hour CPU run into 14~minutes at $\nx=64,\ \nt=256$. Second, the \emph{relative} ranking of the three GPUs reverses between the two grid families: at $\nx=64$ the A100
(GPU-3) is the \emph{slowest} of the three ($\sim$32--45$\times$), while at $\nx=128$ it becomes the \emph{fastest} ($\sim$97--101$\times$). This is explained by memory bandwidth: the A100 provides $\sim$2~TB/s of HBM2e bandwidth versus $\sim$192~GB/s for the RTX~4070 Laptop GPU; at $\nx=64$ the working set is small enough to partly fit in CPU cache and to be handled comparably well by all three GPUs, whereas at $\nx=128$ the four-times larger grid stresses memory bandwidth enough that the A100's advantage fully
materializes. A similar, smaller effect explains the CPU speedup drop at $\nx=64,\ \nt=4096$ ($\sim$50$\times\to$36$\times$): the 16-fold larger temporal tensor pushes the CPU working set beyond its 8~MB L3 cache, while the GPU is comparatively unaffected because its memory latency is hidden by massive thread-level parallelism.}

\subsection{Algorithm convergence}
\label{sec:convergence}

\begin{figure*}
    \centering
    \includegraphics[width=\linewidth]{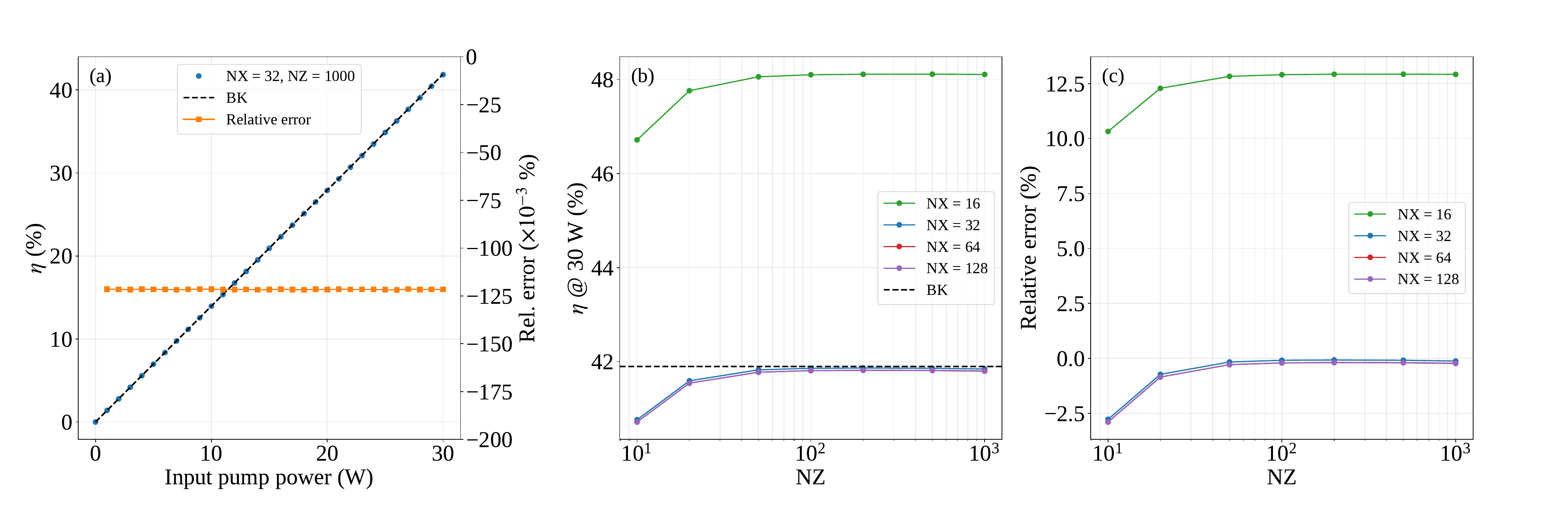}
    \caption{Algorithm convergence analysis for Boyd and Kleinman (BK) analytical solution. Panel (a): linear analytical and numerical power scaling. Panels (b, c): convergence analysis for different grid sizes.}
    \label{fig:convergence}
\end{figure*}

\edited{The convergence analysis was performed using the focused Gaussian beam solution for the second-harmonic generation (SHG) process, as given by Boyd and Kleinman~\citep{boyd1968parametric}. In this framework, under the undepleted pump approximation ($\partial A_p/\partial z = 0$) and assuming negligible absorption ($\alpha_p = \alpha_s = 0$), and for the specific mismatch value $\dk = -3.2/\lcr$, the SHG efficiency reaches its maximum when $\xi = 2.84$. This analysis confirms that the analytical SHG efficiency follows a linear dependence on the pump power~\citep{sanchez2024cuda3D}. Figure~\ref{fig:convergence}(a) shows the SHG efficiency as a function of the CW pump power up to 30~W. As observed, the Boyd–Kleinman (BK) analytical solution and the numerical simulation exhibit a relative error below 0.2\%. The convergence behavior for different grid resolutions is also presented, with varying $\nx = \ny$ and $\nz$, at a fixed pump power of $\Pin = 30$~W (panel (b)), along with the corresponding relative error (panel (c)). The quantitative convergence results are summarized, and a clear grid-resolution recommendation is provided, namely $\nx = \ny = 32$ or $64$.}

\section{Illustrative example}
\label{sec:examples}

\begin{figure*}
    \centering
    \includegraphics[width=\linewidth]{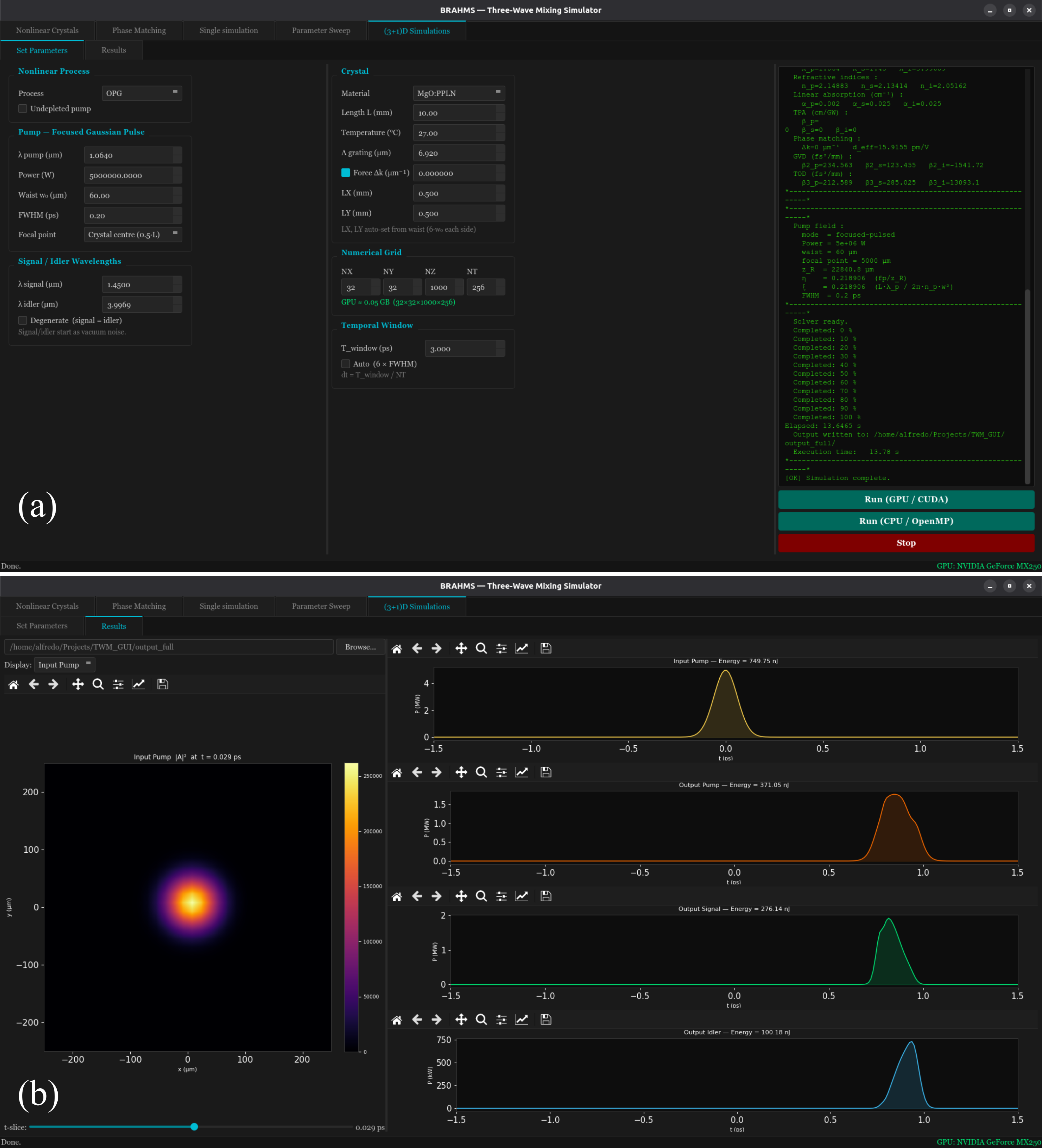}
    \caption{Example of (3+1)D simulations for an OPG nonlinear process in an MgO:PPLN crystal. Panel (a): The tab where simulation parameters are set (pump peak power, wavelengths, beam waist, grid size, crystal length, etc.) and where the CPU or GPU execution mode is selected in the lower right section. Panel (b): Simulation results in the time domain for all electric fields.}
    \label{fig:example_fp}
\end{figure*}

\edited{The project repository~\citep{repo} contains a more comprehensive explanation of how to use the GUI in typical third-order process applications, which are useful for making predictions, estimates, and planning experiments. In this section, a simple example illustrating how to perform a (3+1)D simulation is presented. This type of simulation represents the most general problem that \brahms~can solve and is the core of this work. Figure~\ref{fig:example_fp}(a) shows the \texttt{(3+1)D Simulations/Set Parameters} tab, which contains the parameters for (3+1)D simulations in the \texttt{focused-pulsed} mode. As a concrete example, the OPG process was selected, with wavelengths $\left( \lambda_p, \lambda_s, \lambda_i \right) = \left(1064, 1450, 3997\right)$~nm. The remaining parameters used in the simulation are also visible in this tab, including pump peak power, wavelengths, beam waist, grid size, crystal length, and others. The functional form of the field is identical to Eq.~\ref{eq:gaussianbeamc}, but with additional time dependence at $z=0$, as}
\begin{equation}
    A_p(x,y, \tau) = \frac{A_{p0}}{1-i\xi}\exp\left[-\frac{x^2+y^2}{w_{p}^2(1-i\xi)}-\left(\frac{\tau}{\tau_0}\right)^2\right],
\end{equation}
\edited{where $\tau_0=200$~fs. Once the parameters are set, the user can run the simulation by clicking either the \texttt{Run CPU} or \texttt{Run GPU} button, although using the GPU mode is highly recommended.}

\edited{The corresponding outputs of the three fields are shown in Fig.~\ref{fig:example_fp}(b) in the \texttt{(3+1)D Simulations/Results} tab. As can be seen, the pulse duration is similar to the input pulse, but they differ in amplitude and exhibit a time delay due to dispersive effects. It should be noted that the code calculates the output power by spatially integrating the intensity field according to:}
\begin{equation}
    P_{\lambda}(\tau) =\frac{1}{2}c\varepsilon_0 n_{\lambda} \iint_{\mathcal{S}_{x,y}} \left| A_{\lambda}(x,y,\lcr, \tau) \right|^2 dxdy,    
\end{equation}

\edited{where $P_{\lambda}(\tau)$ is a function of time, as shown in Fig.~\ref{fig:example_fp}(b), and $\mathcal{S}_{x,y}$ is the transversal area of the nonlinear crystal. On the left side of the window, a graph of the beam profile for all involved electric fields is displayed (selectable via the \texttt{Display} menu in the upper right), and users can navigate through the temporal window using the slice slider in the lower left.}

\section{Conclusions}
\label{sec:conclusions}
We have presented~\brahms, \edited{a cross-platform (Linux/Windows) numerical solver, with GPU and CPU (OpenMP) backends,} for the coupled $\chi^{(2)}$ nonlinear Schrödinger equations in full $(3+1)$D, designed to model three-wave mixing processes involving focused and pulsed Gaussian beams. The package provides a unified framework that simultaneously incorporates spatial diffraction, temporal dispersion, \edited{spatial walk-off,} phase mismatch, and nonlinear coupling, overcoming the dimensional and memory limitations that have traditionally restricted numerical treatments of bulk nonlinear interactions to reduced geometries.

The numerical implementation is based on a SSFM, combined with an efficient \edited{GPU- and CPU-oriented} data layout that enables the computation of large numbers of batched 1D- and 2D-FFTs. \edited{On the GPU backend, executing all operations, including memory management and field evolution, on the device allows} realistic (3+1)D simulations of field propagation that would otherwise be computationally demanding on conventional CPU-based approaches\edited{; the equivalent CPU (OpenMP) backend makes the same simulations accessible on machines without a GPU}. Representative benchmarks demonstrate that simulations with large spatial and temporal grids can be performed within practical runtimes on modern GPU architectures\edited{, with speedups greater than $30\times$ over the CPU backend depending on grid size and GPU model}.

\edited{An illustrative example of how to use the GUI to perform a (3+1)D simulation was presented. In this example, GPU usage was recommended to run the \texttt{focused-pulsed} mode because the package is optimized for this type of simulation. The results can be visualized using the visualization capabilities provided by the GUI and are also saved in HDF5 files for subsequent post-processing.}

The modular structure of the~\cpp/CUDA \edited{and \cpp/OpenMP} backend\edited{s}, together with a \edited{cross-platform GUI} and JSON-based configuration, makes~\brahms~accessible to users with minimal programming experience while remaining flexible enough for advanced customization. The package is therefore well suited both as a research tool for investigating complex nonlinear wave-mixing phenomena and as a practical simulation platform for experimental planning in bulk nonlinear optics.

\section*{Acknowledgements}
The authors thank Himani Sharma and Rojalin Padhi for kindly providing access to some of the GPUs used in this work.

\section*{Declaration of generative AI and AI-assisted technologies in the manuscript preparation process}

During the preparation of this work, the author(s) used \texttt{Claude Code} for optimizing the developed package. The author(s) reviewed and edited the output as needed and take full responsibility for the content of the published article. 

\bibliographystyle{elsarticle-num}
\bibliography{biblio}

\end{document}